\documentclass[prl,twocolumn,showpacs]{revtex4}

\usepackage{graphicx}
\usepackage{amsmath}

\begin{document}

\title{Transition from a strongly interacting 1D superfluid to a Mott insulator}

\author{Thilo St\"{o}ferle, Henning Moritz, Christian Schori, Michael K\"{o}hl$^\dag$ and Tilman Esslinger}

\affiliation{Institute of Quantum Electronics, ETH Z\"{u}rich
H\"{o}nggerberg, CH--8093 Z\"{u}rich, Switzerland}

\date{October 16, 2003}

\begin{abstract}

We study one-dimensional trapped Bose gases in the strongly
interacting regime. The systems are created in an optical lattice
and are subject to a longitudinal periodic potential. Bragg
spectroscopy enables us to investigate the excitation spectrum of
the one-dimensional gas in different regimes. In the superfluid
phase a broad continuum of excitations is observed which calls for
an interpretation beyond the Bogoliubov spectrum taking into
account the effect of quantum depletion. In the Mott insulating
phase a discrete spectrum is measured. The excitation spectra of
both phases are compared to the three-dimensional situation and to
the crossover regime from one to three dimensions. The coherence
length and coherent fraction of the gas in all configurations are
measured quantitatively. We observe signatures for increased
fluctuations which are characteristic for 1D systems. Furthermore,
ceasing collective oscillations near the transition to the Mott
insulator phase are found.

\end{abstract}

\pacs{05.30.Jp, 03.75.Kk, 03.75.Lm, 73.43.Nq}

\maketitle

Quantum gases trapped in the periodic potential of an optical
lattice have opened a new experimental window on many-particle
quantum physics. The recent observation of the quantum phase
transition from a superfluid to a Mott insulating phase in a Bose
gas \cite{Greiner2002a} has offered a first glimpse on the physics
which is now becoming experimentally accessible. However, the full
wealth of possibilities has yet to be explored. Besides
controlling the effect of interactions in the trapped gas, it is
conceivable to induce disorder, to change the dimensionality of
the system, or to trap Fermi gases or Bose-Fermi mixtures. The
realization of these systems is expected to provide a deeper
understanding of general concepts related to superfluidity and
superconductivity.

Here we use the optical lattice to realize a strongly interacting
Bose gas in one spatial dimension and to study the crossover to
three dimensions. Emphasis is put on the measurement of excitation
spectra which characterize the transition from the superfluid
\cite{Menotti2003,Rey2003} to the Mott insulating state
\cite{Sheshadri1993,Greiner2002a,Sachdev2002}. Several features
observed in the spectra go beyond the description of current
theoretical models.

Degenerate Bose gases trapped in the lowest band of an optical
lattice can be modelled using the Bose-Hubbard Hamiltonian
\cite{Fisher1989,Freericks1994,Kuehner1998,Jaksch1998}, in which
the hopping of atoms between neighboring lattice sites is
characterized by the tunnelling matrix element $J$, while the
interaction energy for two atoms occupying the same site is given
by $U$. The physics of this model is governed by the ratio between
$U$ and $J$, i.e$.$ between interaction and kinetic energy. This
parameter can be controlled by changing the depth of the lattice
potential. If the ratio $U/J$ is below a critical value the atoms
are superfluid. Above this value the system becomes Mott
insulating. We access the one-dimensional regime
\cite{Fisher1989,Buechler2003,Batrouni2002} using an anisotropic
optical lattice consisting of three mutually perpendicular
standing waves. By choosing large potential depths in two axes we
can selectively suppress tunnelling and hopping is possible only
along one dimension, so that an array of one-dimensional tubes
with periodic modulation along their axis is formed.

We produce almost pure Bose-Einstein condensates of typically
$1.5\times 10^5$ $^{87}$Rb atoms in the $|F=2, m_F=2\rangle$
hyperfine ground state in a magnetic trap with trapping
frequencies of $\omega_x=2 \pi \times 18\,\text{Hz}$, $\omega_y=2
\pi \times 20\,\text{Hz}$, and $\omega_z=2 \pi \times
22\,\text{Hz}$. The optical lattice is formed by three
retro-reflected laser beams which are derived from laser diodes at
a wavelength of $\lambda=826$\,nm \cite{Moritz2003}. At the
position of the condensate the beams are circularly focused to
$1/e^2$-radii of $120$\,$\mu\text{m}$ ($x$ and $y$ axes) and
$105$\,$\mu\text{m}$ ($z$). The three beams possess mutually
orthogonal polarizations and their frequencies are offset with
respect to each other by several ten $\text{MHz}$. The linewidth
of the lasers is of the order of $10\,\text{kHz}$.

In order to load the condensate into the ground state of the
optical lattice, the intensities of the lasers are slowly
increased to their final values using an exponential ramp with a
time constant of $25\,\text{ms}$ and a duration of
$100\,\text{ms}$. The resulting optical potential depths
$V_{x,y,z}$ are proportional to the laser intensities and are
conveniently expressed in terms of the recoil energy
$E_R=\frac{\hbar^2 k^2}{2 m}$ with $k=\frac{2 \pi}{\lambda}$ and
the atomic mass $m$. To prepare an array of one-dimensional tubes,
two lattice axes are ramped to $V_{\bot}\equiv V_x=V_z=30\,E_R$
and the third one to a much lower value $V_{ax,0}\equiv V_y$. In
this configuration the transverse tunnelling rates $J_x$ and $J_z$
are small and contribute a correction of the order of
$J_{x,z}/\mu\ll 1$ to the 1D characteristics of the individual
tubes, where $\mu$ is the chemical potential of the sample. It is
convenient to include the anisotropic tunnelling between all
nearest neighbor sites which yields the resulting
$J=2(J_x+J_y+J_z)$.

We study the excitation spectrum by employing amplitude modulation
of the axial lattice potential $V_{ax}$ to perform two-photon
Bragg spectroscopy \cite{Stenger1999}. The lattice potential takes
the form $V_{ax}(y,t) = (V_{ax,0} + A_{mod}\sin{(2\pi\nu_{mod}t)})
\sin^2(ky)$. The modulation with amplitude $A_{mod}$ and frequency
$\nu_{mod}$ introduces two sidebands with frequencies $\pm
\nu_{mod}$ relative to the lattice laser frequency which define
the energy $h\nu_{mod}$ of the excitation. Due to the Bragg
condition atoms scattering two photons receive a momentum transfer
of $0\hbar k$ or $2\hbar k$. In contrast to applying a potential
gradient across the lattice \cite{Greiner2002a}, this method is
not susceptible to effects like Bloch oscillations and Zener
tunnelling which occur for low axial lattice depths. Furthermore,
the excitation energy is precisely determined and does not involve
any parameters that need calibration.

After the excitation, the experimental sequence continues by
ramping down the lattice potentials linearly in 15\,ms to
$V_{ax}=V_{\bot}=4\,E_R$ where the atoms are able to tunnel again
in all three dimensions between the sites of the lattice. To allow
for re-thermalization of the system, the atoms are kept at this
lattice depth for $5\,\text{ms}$. Then all optical and magnetic
potentials are suddenly switched off. The resulting matter wave
interference pattern is detected by absorption imaging after
$25\,\text{ms}$ of ballistic expansion. The width of the central
momentum peak is taken as a measure of how much energy has been
deposited in the sample by the excitation. If the energy increase
is small, the peak is well fitted by a bimodal distribution. For
resonant excitation there is only a single gaussian component,
reflecting that the temperature of the atoms has significantly
increased. To be independent of the shape of the peak we use the
full width at half maximum (FWHM) as a measure of the introduced
energy. Although this underestimates small energy increases, the
important resonances and features of the spectra are well shown.

The duration $t_{mod}=30\,\text{ms}$ and amplitude $A_{mod}=0.2
V_{ax,0}$ of the modulation are chosen such that the resulting
excitation of the condensate does not exhibit saturation effects
for all measurements presented here. We have verified that all
atoms remain in the lowest Bloch band by adiabatically switching
off the lattice potentials \cite{Greiner2001b} after the
modulation. When we load a cold thermal cloud into the lowest
Bloch band and apply our modulation scheme, we do not observe
excitations.

\begin{figure}[htbp]
  \includegraphics[width=.6\columnwidth,clip=true]{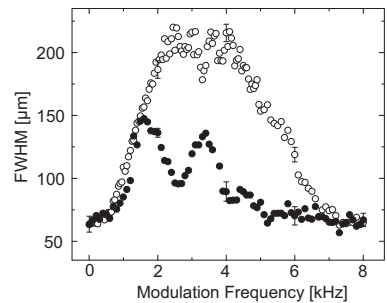}
  \caption{Spectroscopy of the 1D superfluid (open circles) and the Mott
  insulating phase (filled circles) with values of $U/J$ of approximately 2.3 and 14
  respectively. Error bars reflect the statistical error of 5~measurements.}
  \label{fig1}
\end{figure}

Figure~\ref{fig1} displays the fundamental change in the
excitation spectrum for a 1D Bose gas when the crossover from the
superfluid to the Mott insulating phase occurs: The broad
continuum of the superfluid contrasts with the discrete spectrum
of the Mott insulator. One surprising feature is that we can
excite the superfluid with our scheme at large $h\nu_{mod}$
contrary to predictions for the weakly interacting superfluid in
an optical lattice formed by a single standing wave
\cite{Menotti2003}. In our experiment strong interactions lead to
a significant quantum depletion, which is $\approx 50\%$ for the
1D configuration even with $U/J=2.3$ \cite{Kraemer2003}. Therefore
this parameter may not be regarded small as in standard Bogoliubov
theory, but higher order excitations should be taken into account
\cite{Hugenholtz1959}. In combination with the broken
translational invariance in the inhomogeneous trap, this could
explain the non-vanishing excitation probability observed in the
experiment at high energies \cite{BuechlerTBP}.

\begin{figure}[htbp]
  \includegraphics[width=0.84\columnwidth,clip=true]{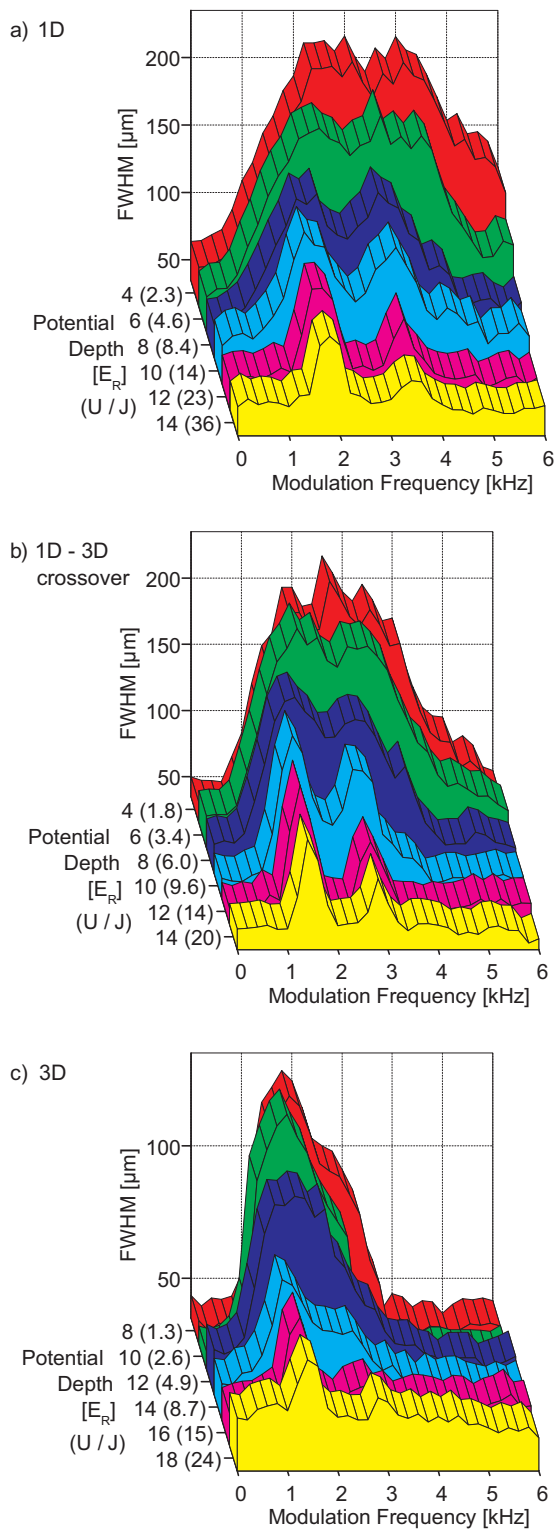}
  \caption{The measured excitation spectrum of an array of 1D gases ($V_{\bot}=30\,E_R$)
  is shown in (a) for different values of $V_{ax,0}$. The interaction ratios
  $U/J$ given in brackets are calculated numerically using a band structure model in the tight-binding
  approximation \cite{Jaksch1998}. Spectrum (c) shows the superfluid to Mott insulator transition in the 3D case
  ($V_\perp=V_{ax,0}$). The crossover region between the one- and the three-dimensional
  system ($V_{\bot}=20\,E_R$) is shown in (b).}
  \label{fig2}
\end{figure}

A full series of spectra for different values of $U/J$ ranging
from the superfluid via the crossover region to the Mott
insulating phase is shown in figure~\ref{fig2}. Figure~\ref{fig2}a
displays the 1D situation with $V_\perp=30\,E_R$. The crossover
from one to three dimensions (figure~\ref{fig2}b) is achieved by
reducing the transverse confinement of the optical lattice to
$V_{\bot}=20\,E_R$. Then the tunnelling time between the 1D tubes
is of the order of $t_{mod}$ and thus the interaction between the
tubes is not negligible any more \cite{Giamarchi2003}. Finally,
figure~\ref{fig2}c shows the three-dimensional case with
$V_\perp=V_{ax,0}$.

In one dimension (figure~\ref{fig2}a) we observe the appearance of
the discrete structure, which is characteristic for the Mott
insulating phase, between $U/J= 4$ and $U/J= 8$. Above $U/J= 20$
there is no more background due to the vanishing superfluid
component. Our results are in accordance with the prediction $U/J=
5.8$ for entering the $n=1$ Mott insulator based on a mean-field
theory \cite{Fisher1989,Jaksch1998}. Calculations beyond the
mean-field approach give an onset of the Mott insulating phase in
the homogeneous 1D system at $U/J\approx 1.8$ \cite{Kuehner1998}.
However, the finite size of the trap prohibits a sharp transition
\cite{Batrouni2002}, so that the fraction of Mott insulating atoms
increases gradually with $U/J$.

For the superfluid we obtain spectra which differ significantly
from the results of ref.~\cite{Greiner2002a}, since the superfluid
excitations decrease at higher energies. This decrease is rather
slow for the 1D gas but becomes more pronounced when the
tunnelling between the 1D gases is increased (Figure~\ref{fig2}b
and Figure~\ref{fig2}c). Our excitation scheme does not induce
dephasing that occurs when the strongly interacting condensate is
accelerated near the edge of the Brillouin zone
\cite{Bronski2001}. This might cause the broadening and the
background in the tilted lattice experiments at high energies in
ref.~\cite{Greiner2002a}. The width of the superfluid spectra for
the 1D gas is on the same order as twice the width of the lowest
band for Bogoliubov excitations \cite{Kraemer2003b}.

\begin{figure}[htbp]
  \includegraphics[width=.7\columnwidth,clip=true]{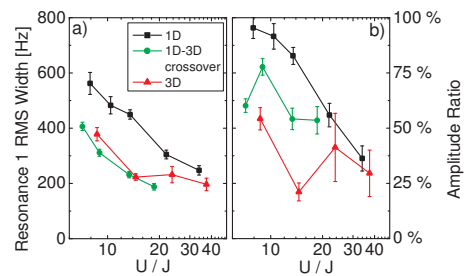}
  \caption{(a) Width of the first resonance peak in the spectrum of the Mott insulator.
  (b) Ratio of the amplitudes of the second and the first peak of the spectrum of the Mott insulator.
  The error bars mark the error of the gaussian fits.}
  \label{fig3}
\end{figure}

In the Mott insulating phase we find the first resonant peak for
all data sets close to the calculated value of $U$. A second peak
appears at $(1.91\pm 0.04)$ times the energy of the first
resonance, somewhat smaller than the value of 2 reported in
\cite{Greiner2002a}. This resonance might be attributed to defects
where lattice sites with $n=1$ atom next to sites with $n=2$ atoms
are being excited. For the 1D system and in the dimensional
crossover regime (Fig.~\ref{fig2}a and b) a much weaker resonance
appears at $(2.60\pm 0.05)$ times the energy of the first
resonance which could indicate higher order processes of two atoms
tunnelling simultaneously. In figure~\ref{fig3}a we plot the rms
width of the first resonance in the excitation spectrum of the
Mott insulating phase when fitted by a gaussian. In
figure~\ref{fig3}b we show the ratio of the amplitudes of the
second and the first peak. Apparently, in the 1D system the first
peak is wider and the second peak more pronounced as compared to
the 3D situation, which could be an indication of increased
fluctuations in 1D systems.

\begin{figure}[htbp]
  \includegraphics[width=.7\columnwidth,clip=true]{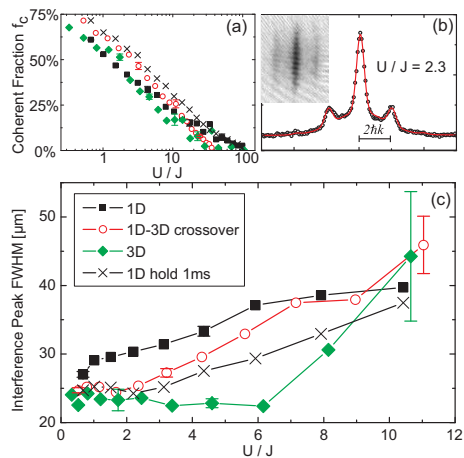}
  \caption{(a) Coherent fraction vs. $U/J$. The error bars are determined by the statistical error of 4~measurements.
  (b) The column sum of the optical density (circles)
  and the fits (solid line) from which the number of coherent and incoherent atoms
  are deduced for the 1D case. The inset shows the absorption image after $10\,\text{ms}$ of time-of-flight
  (dimensions $360\,\mu\text{m}\times 467\,\mu\text{m}$). (c) Width of the central momentum peak vs.
  $U/J$. }
  \label{fig4}
\end{figure}

Compared to the superfluid properties the coherence properties of
the system provide complementary information about the state of
the gas. They are probed by studying the matter wave interference
pattern \cite{Orzel2001,Greiner2002a}. Here we first prepare the
array of 1D systems as above but do not apply our excitation
scheme. Instead, after holding the atoms at the final lattice
depth for $t_h=30\,\text{ms}$, we increase $V_{ax}$ rapidly
($<40\,\mu\text{s}$) to about $25\,E_R$ and then abruptly switch
off all optical and magnetic trapping potentials. This procedure
projects the different initial configurations onto the same Bloch
state. To extract the number of coherent atoms $N_{coh}$ from the
interference pattern, the peaks \cite{remarkInterference} at
$0\hbar k$, $\pm 2\hbar k$ and $\pm 4\hbar k$ are fitted by
gaussians (Fig.~\ref{fig4}b). Incoherent atoms give rise to a
broad gaussian background which dominates for higher $V_{ax,0}$.
Taking this fit as a measure of the number of incoherent atoms
$N_{incoh}$, we calculate the coherent fraction
$f_c=\frac{N_{coh}}{N_{coh}+N_{incoh}}$. As shown in
figure~\ref{fig4}a, $f_c$ decreases slowly to zero for increasing
values of $U/J$ and appears almost independent of the
dimensionality. This coincides with the prediction that for
strongly interacting Bose gases in optical lattices the superfluid
fraction can be significantly different from the coherent
fraction, and that the decrease of $f_c$ is not a sufficient
signature of entering the Mott insulating phase \cite{Roth2003}.
In figure~\ref{fig4}c we plot the width of the central peak of the
interference pattern, which is a measure of the coherence length
of the gas. An increasing width is a good indicator for the
presence of a Mott insulating phase since even a small Mott
insulating domain reduces the coherence length of the sample, as
elucidated in numerical calculations \cite{Kollath2003}. Our data
shows that the increase in width starts at much lower values of
$U/J$ for the 1D gas than for the 3D gas. This supports the
expectation that due to the more pronounced quantum fluctuations
in the 1D geometry the gas enters the Mott insulating state at
lower values of $U/J$
\cite{Kuehner1998,Batrouni2002,Buechler2003,Kollath2003}.
Experimentally, thermal fluctuations may also contribute to the
observed width. However, by comparing the width of the
interference peak for different hold times ($t_h=1$\,ms and
$t_h=30$\,ms) we find that the primary effect of the additional
heating is an overall increase of the width rather than a change
of the slope of the curve.

One feature of the Mott insulating phase is its lack of
compressibility \cite{Fisher1989}. If a potential gradient is
applied across the lattice, the inability to redistribute atoms in
the lattice results in a breakdown of the flow of atoms. Even if
only a fraction of the atoms in our inhomogeneous trap is locally
incompressible and pinned in the Mott insulating state
\cite{Batrouni2002}, collective oscillations should suspend. In
the experiment, the array of one-dimensional gases is prepared in
the same way as before. Then, we apply a small potential gradient
(far less than any resonance gradients \cite{Greiner2002a}) for
$1\,\text{ms}$ to accelerate the atoms in the axial direction. We
observe a very strong increase in the damping of the small
amplitude dipole oscillations even for very low $V_{ax,0}$, a
behaviour which is different from the three-dimensional gas.
Critical damping is reached at $V_{ax,0}=3\,E_R$, and at
$V_{ax,0}=10\,E_R$ we can not displace the atom cloud any more.

In conclusion, we have prepared an array of one-dimensional
strongly interacting Bose gases and measured the excitation
spectra in the superfluid and the Mott insulating regime. The
spectrum of the superfluid exhibits excitations at high energies
which are not predicted by current theories for weakly interacting
bosons in an optical lattice and have so far only been discussed
in the context of superfluid Helium. This shows that the strong
interactions and the significant quantum depletion change the
properties of the Bose gas considerably, already much below the
crossover to the Mott insulator. The effect of reduced
dimensionality on the crossover from superfluid to Mott insulator
was most pronounced in the measurements of the coherence length of
the gas.

We would like to thank H.~P.~B{\"u}chler and G.~Blatter for
insightful discussions, and SNF and QSIT for funding.


\begin{thebibliography}{}

\bibitem[\dag]{email}{email:Koehl@iqe.phys.ethz.ch}


\bibitem{Greiner2002a}M.~Greiner {\it et al.}, Nature~\textbf{415}, 39 (2002).

\bibitem{Menotti2003}C.~Menotti, M. Kr{\"a}mer, L.Pitaevskii, S. Stringari, Phys. Rev.~A~\textbf{67}, 053609 (2003).

\bibitem{Rey2003}A.~M.~Rey \textit{et al.}, J. Phys.~B~\textbf{36}, 825 (2003).

\bibitem{Sheshadri1993}K. Sheshadri, H.R. Krishnamurthy, R. Pandit ,T.V. Ramakrishnan, Europhys. Lett.~\textbf{22}, 257 (1993).

\bibitem{Sachdev2002}S.~Sachdev, K. Sengupta, S.M. Girvin, Phys. Rev.~B~\textbf{66}, 075128 (2002).

\bibitem{Fisher1989}M.~P.~A.~Fischer, P.B. Weichmann, G. Grinstein, D.S. Fisher, Phys. Rev.~B~\textbf{40}, 546 (1989).

\bibitem{Freericks1994}J.~K.~Freericks, H.~Monien, Europhys. Lett.~\textbf{26}, 545
(1994).

\bibitem{Kuehner1998}T.~D.~K\"{u}hner, H.~Monien, Phys. Rev.~B~\textbf{58},
14741 (1998).

\bibitem{Jaksch1998}D.~Jaksch {\it et al.}, Phys. Rev. Lett.~\textbf{81}, 3108 (1998).

\bibitem{Batrouni2002}G.~G.~Batrouni \textit{et al.}, Phys. Rev. Lett.~\textbf{89}, 117203 (2002).


\bibitem{Buechler2003}H.~P.~B\"{u}chler, G. Blatter, W. Zwerger, Phys. Rev. Lett.~\textbf{90}, 130401 (2003).

\bibitem{Moritz2003}H.~Moritz, T. St{\"o}ferle, M. K{\"o}hl, T. Esslinger, e-print cond-mat/0307607.

\bibitem{Stenger1999}J.~Stenger {\it et al.}, Phys. Rev. Lett. {\bf
82}, 4569 (1999).

\bibitem{Greiner2001b}M.~Greiner {\it et al.}, Phys. Rev. Lett.~\textbf{87}, 160405 (2001).

\bibitem{Kraemer2003} M. Kr{\"a}mer, C. Menotti, L. Pitaevskii, S. Stringari, Eur. Phys. J. D {\bf
27}, 247 (2003).

\bibitem{Hugenholtz1959}N.M. Hugenholtz, D. Pines, Phys. Rev.~\textbf{116}, 489 (1959).

\bibitem{BuechlerTBP}H.~P.~B\"{u}chler, private communication.

\bibitem {Giamarchi2003} A. F. Ho, M. A. Cazalilla, T.Giamarchi, e-print cond-mat/0310382.

\bibitem{Bronski2001}J.~C.~Bronski, L.D. Carr, B. Deconinck, J.N. Kutz, Phys. Rev. Lett.~\textbf{86}, 1402 (2001).

\bibitem{Kraemer2003b} M. Kr{\"a}mer, private communication.

\bibitem{Orzel2001}C.~Orzel {\it et al.}, Science~\textbf{291}, 2386 (2001).
\bibitem{Roth2003}R.~Roth, K.~Burnett, Phys. Rev.~A~\textbf{67}, 031602 (2003).

\bibitem{remarkInterference}At $V_{\bot}=30\,E_R$, coherence between the individual 1D systems is lost after a few milliseconds
\cite{Moritz2003}. Perpendicular to the axis of the tubes the
expansion of the ground state is gaussian.
%

\bibitem{Kollath2003} C.~Kollath, U. Schollw{\"o}ck, J. von Delft, W. Zwerger, e-print
cond-mat/0310388.


\end{thebibliography}
\end{document}